\documentclass[11pt]{amsart}

\usepackage[utf8]{inputenc}
\usepackage[T1]{fontenc}
\usepackage[english]{babel}
\usepackage{amsmath,amssymb,amsthm}
\usepackage{mathrsfs}
\usepackage{hyperref}

\hypersetup{
  colorlinks=true,
  linkcolor=blue,
  citecolor=blue,
  urlcolor=blue
}

\theoremstyle{definition}
\newtheorem{definition}{Definition}[section]
\theoremstyle{plain}

\newtheorem{proposition}[definition]{Proposition}
\theoremstyle{remark}
\newtheorem{remark}[definition]{Remark}

\numberwithin{equation}{section}

\newcommand{\C}{\mathbb{C}}

\newcommand{\A}{\mathcal{A}}
\newcommand{\Hc}{\mathbb{H}}
\newcommand{\Oc}{\mathbb{O}}
\newcommand{\Cbb}{\mathbb{C}}
\newcommand{\Rbb}{\mathbb{R}}

\begin{document}

\title[SU(3)-covariant noncommutative KP hierarchy]{A Lorentzian SU(3)-covariant noncommutative KP hierarchy and hypercomplex gauge fields}

\author{Jean-Pierre Magnot}
\address{$^*$  Univ. Angers, CNRS, LAREMA, SFR MATHSTIC, F-49000 Angers, France ;\\ 
Lepage Research Institute, 17 novembra 1, 081 16 Presov, Slovakia;
	\\ and \\  Lyc\'ee Jeanne d'Arc, \\ Avenue de Grande Bretagne, \\ 63000 Clermont-Ferrand, France}
    \email{jean-pierr.magnot@ac-clermont.fr}
\date{}

\begin{abstract}
We propose a formal framework for a noncommutative Kadomtsev--Petviashvili (KP) hierarchy which is covariant under the action of $SU(3)$ and compatible with a Lorentzian structure encoded in a twisted quaternionic (or Clifford) algebra. The starting point is a formal pseudodifferential operator $L$ built from an abstract derivation $D$ of Dirac type and coefficients in an associative algebra $\A$ that combines spin degrees of freedom (twisted quaternions, Clifford algebras) and color degrees of freedom (an internal $SU(3)$ factor, possibly realized via the octonions). In this way we obtain a hierarchy of formal partial differential equations which are Lorentz invariant and $SU(3)$ covariant and can be interpreted as integrable sectors of nonabelian gauge theories in $(3+1)$ dimensions and of their dimensional reductions.
\end{abstract}

%PACS: 02.30.Ik, 02.30.Jr, 02.40.Yy, 11.10.Lm, 11.15.-q


\maketitle

\section{Introduction}

Integrable hierarchies of Kadomtsev--Petviashvili (KP) type and their reductions (KdV, NLS, Boussinesq, etc.) play a central role in the theory of solitons and in the development of the inverse scattering transform \cite{AblowitzSegur1981,Dickey2003}. Since the early work on Gelfand--Dickey hierarchies and Lax formulations, it has become clear that many integrable equations in $(1+1)$ and $(2+1)$ dimensions can be obtained as reductions of more universal systems, in particular of self-dual Yang--Mills (SDYM) equations \cite{AblowitzChakravartyTakhtajan1993,AblowitzChakravartyHalburd2003,Ward1985,MasonWoodhouse1996}.

In parallel, a wide range of noncommutative generalizations of classical integrable hierarchies has been developed, where the coefficients of the pseudodifferential operators take values in a noncommutative associative algebra. We refer in particular to the work of Dimakis and M\"uller-Hoissen on the noncommutative KP hierarchy and its extensions \cite{DimakisMuellerHoissen2005,DimakisMuellerHoissen2007} and to the literature on noncommutative integrable systems and the noncommutative Ward conjecture \cite{Hamanaka2006Ward,NoncommutativeIntegrableHamanaka}. In these settings, noncommutativity may model a Moyal-type noncommutative geometry or internal matrix degrees of freedom.

The present work is in this line, but with a more geometric and “physical” motivation: to construct a \emph{formal} KP-type hierarchy in which
\begin{itemize}
  \item the basic derivation $D$ is of Dirac type, encoding a Lorentzian structure of signature $(1,3)$ through a twisted quaternionic or suitable Clifford algebra;
  \item the coefficient algebra $\A$ combines a ``space-time/spin'' factor (twisted quaternions, Clifford algebras) and an internal color factor (an algebra associated with $SU(3)$, possibly through an octonionic realization);
  \item the resulting hierarchy is \emph{covariant} under the action of $SU(3)$ by conjugation on the coefficients, and \emph{compatible} with the spinorial action of the Lorentz group on the quaternionic/Clifford part.
\end{itemize}

From the viewpoint of high energy and gauge theory, our motivation is to isolate and describe \emph{integrable sectors} of nonabelian gauge/Dirac systems with internal $SU(3)$ symmetry. In four space-time dimensions, Yang--Mills theories with gauge group $SU(3)$ provide the classical field-theoretic backbone of quantum chromodynamics, while Dirac operators encode spin-$\tfrac{1}{2}$ matter fields. The full classical dynamics is highly nonlinear and, in general, nonintegrable. However, it has long been understood that self-dual or generalized self-dual configurations of the field strength define special sectors with enhanced symmetry and integrable structure, as in the self-dual Yang--Mills (SDYM) reductions leading to KdV-, KP- or NLS-type equations. The present work is motivated by the idea that a similar phenomenon may occur in a more elaborate hypercomplex gauge setting, where Lorentz geometry is encoded in a twisted quaternionic (or Clifford) algebra, colour degrees of freedom are carried by an $SU(3)$ factor, and the evolution parameters themselves are promoted to noncommuting quaternionic ``times''. In this framework, the generalized KP hierarchy built from a Dirac-type derivation $D$ may be interpreted as an effective description of certain constrained, yet infinitely rich, families of $SU(3)$ gauge/Dirac fields. The resulting hypercomplex, $SU(3)$-covariant integrable equations can then be viewed as candidates for integrable subsectors or toy models of nonabelian flux-tube configurations, colour solitons, or reduced SDYM-like geometries, where the interplay between Lorentz symmetry, internal $SU(3)$ symmetry and noncommutative time structure is made explicit and tractable.
In this sense, the present paper should be regarded as providing a formal integrable framework tailored to such hypercomplex gauge configurations, rather than a fully developed field-theoretic model. We now turn to the algebraic and geometric ingredients underlying this construction.

On the purely formal side, we start from a differential algebra $(\A,D)$, where $\A$ is a (possibly noncommutative) associative complex algebra and $D$ is an abstract derivation. We then consider the algebra of formal pseudodifferential operators $\Psi(\A,D)$ consisting of formal series with integer powers (positive and negative) of $D$ and coefficients in $\A$, finite towards $+\infty$. The fundamental commutation relation
\[
  D a = a D + D(a),\qquad a\in\A,
\]
allows one to define an associative algebra structure on $\Psi(\A,D)$ via generalized binomial formulas.

A KP-type Lax operator is then a formal element
\[
  L = D + \sum_{k\ge 1} U_k D^{-k},\qquad U_k\in\A,
\]
and for each $n\ge 1$ we define the $n$-th flow of the hierarchy by the Lax equation
\[
  \partial_{t_n} L = [B_n,L],\qquad B_n=(L^n)_+,
\]
where $(\cdot)_+$ denotes the differential part (nonnegative powers of $D$) of a pseudodifferential operator. As in the classical commutative case, one shows that these flows commute and that the operators
\[
  \mathcal{D}_n := \partial_{t_n} - B_n
\]
satisfy the Zakharov--Shabat compatibility conditions
\[
  [\mathcal{D}_n,\mathcal{D}_m]=0,\qquad \forall n,m\ge 1.
\]
This defines a formal generalized KP hierarchy on $(\A,D)$, without any analytic assumptions.

The new feature in the present framework is the specific choice of the algebra $\A$ and the derivation $D$:
\begin{itemize}
  \item the ``geometric'' part of $\A$ is inspired by the description of Dirac spinors and Maxwell/Dirac equations in Clifford algebra formalisms \cite{Hestenes1966,HestenesSobczyk1984,RodriguesOliveira2007} or, in a more minimal fashion, by a Lorentzianly twisted quaternionic algebra;
  \item the ``internal'' part of $\A$ is built from an algebra realizing the action of $SU(3)$, e.g. a matrix factor $M_3(\C)$ or an octonionic realization in which $SU(3)$ appears as a stabilizer subgroup in the automorphism group $G_2$ \cite{Baez2002,ConwaySmith2003,GunaydinGursey1973};
  \item the derivation $D$ is interpreted as a (formal) Dirac operator, possibly coupled to an $SU(3)$ gauge connection in the spirit of SDYM-based constructions and their integrable reductions \cite{AblowitzChakravartyTakhtajan1993,AblowitzChakravartyHalburd2003,Hamanaka2006Ward}.
\end{itemize}

In this context, the coefficients $U_k$ of $L$ are formal fields with values in $\A$, and the equations coming from the Lax flows provide a family of nonlinear formal PDEs which:
\begin{itemize}
  \item are constructed by a noncommutative KP scheme, in the sense of \cite{DimakisMuellerHoissen2005,DimakisMuellerHoissen2007};
  \item are compatible with a Lorentzian structure encoded in the quaternionic/Clifford factor;
  \item are covariant under $SU(3)$, viewed as a gauge group acting by automorphisms on the internal factor of $\A$.
\end{itemize}

From a physical viewpoint, these hierarchies may be interpreted as \emph{integrable sectors} of nonabelian gauge theories (Yang--Mills with gauge group $SU(3)$) coupled to Dirac fermions, or as relativistic toy models for the study of colored solitonic configurations, flux tubes and integrable gauge backgrounds. They may also be viewed as prototypes of integrable hierarchies in a noncommutative geometric framework à la Connes, where the Dirac operator and the internal algebra are enriched by a hierarchical pseudodifferential structure.

The aim of this work is therefore twofold:
\begin{enumerate}
  \item to present a precise formal construction of a noncommutative KP hierarchy on differential algebras $(\A,D)$ built from hypercomplex structures associated with $\Hc$, $\Oc$ and $SU(3)$;
  \item to discuss possible geometric and physical interpretations of these models, in particular as integrable reductions or deformations of self-dual Yang--Mills-type systems.
\end{enumerate}

\bigskip

\section{Preliminaries: hypercomplex structures, $SU(3)$ and Dirac-type derivations}

In this section we recall the algebraic structures that will be used throughout the paper. Our basic building blocks are hypercomplex algebras (quaternions and octonions), the realization of $SU(3)$ as an internal symmetry group, and Dirac-type derivations on associative algebras built from these ingredients. No evolutionary or ``time'' structure is introduced at this stage; the emphasis is on the purely algebraic and geometric framework.

\subsection{Quaternions and Lorentzian twists}

\subsubsection{The quaternion algebra}

Let $\Hc$ denote the real quaternion algebra, generated by $1,i,j,k$ with relations
\[
  i^2 = j^2 = k^2 = ijk = -1,\qquad
  ij = k,\quad jk = i,\quad ki = j,
\]
and $ji=-k$, etc. Every quaternion can be written uniquely as
\[
  q = a + b i + c j + d k,\qquad a,b,c,d\in\Rbb.
\]
We set $\overline{q} := a - bi - cj - dk$ and define the norm
\[
  N(q) := q\overline{q} = a^2 + b^2 + c^2 + d^2.
\]
The units of norm one form a group isomorphic to $SU(2)$.

Left multiplication by $q$ is represented by a real $4\times4$ matrix $L(q)$ acting on the column vector $(a,b,c,d)^{\mathsf T}$. In the basis $(1,i,j,k)$ one finds
\[
  L(q) =
  \begin{pmatrix}
    a & -b & -c & -d\\
    b &  a & -d &  c\\
    c &  d &  a & -b\\
    d & -c &  b &  a
  \end{pmatrix}.
\]
This realization will serve as a reference point for the Lorentzian twist considered below.

\subsubsection{A Lorentzianly twisted quaternionic structure}

To encode a Lorentzian metric of signature $(1,3)$, we introduce the Minkowski matrix
\[
  M := \mathrm{diag}(1,-1,-1,-1).
\]
Given $q = a + v$ with scalar part $a\in\Rbb$ and purely imaginary part $v = b i + c j + d k$, we define the twisted left multiplication matrix
\begin{equation}\label{eq:twisted-L}
  T(q) := \mathrm{diag}(a,a,a,a) - M\,L(v).
\end{equation}
A direct computation shows that
\[
  T(q) =
  \begin{pmatrix}
    a &  b &  c &  d\\
    b &  a & -d &  c\\
    c &  d &  a & -b\\
    d & -c &  b &  a
  \end{pmatrix}.
\]
This construction can be regarded as defining a new (twisted) product on $\Hc$ by declaring that the matrix $T(q)$ represents left multiplication by $q$ with respect to a basis adapted to the Lorentzian quadratic form $M$. After complexification, this twisted quaternionic algebra can be identified with a model of the even part of the Clifford algebra $Cl_{1,3}$, and thus with the spin group $\mathrm{Spin}(1,3)$ acting on spinorial degrees of freedom.

\subsubsection{Complexified and Clifford-realized quaternions}

We denote by $\Hc^\C := \Hc \otimes_\Rbb \Cbb$ the complexified quaternion algebra. It is well known that
\[
  \Hc^\C \cong M_2(\Cbb),
\]
and that, upon a suitable identification, $\Hc^\C$ realizes the even Clifford algebra $Cl_{1,3}^+$. In particular, Dirac gamma matrices can be expressed in terms of quaternionic (or biquaternionic) structures, and Lorentz transformations act by (complex) quaternionic conjugation. We shall not fix a specific representation; rather, we regard $\Hc^\C$ as an abstract algebra carrying a distinguished class of automorphisms corresponding to Lorentz transformations.

\subsection{Octonions and the internal $SU(3)$ symmetry}

\subsubsection{The octonion algebra and its automorphisms}

The (real) octonion algebra $\Oc$ is an $8$-dimensional nonassociative but alternative algebra over $\Rbb$, with basis
\[
  1, e_1,e_2,e_3,e_4,e_5,e_6,e_7,
\]
and a multiplication encoded by the Fano plane. It is endowed with a positive definite quadratic form
\[
  \|x\|^2 = x\overline{x},\qquad x\in\Oc,
\]
where $\overline{x}$ denotes the standard conjugation. The group of algebra automorphisms $\mathrm{Aut}(\Oc)$ is the compact exceptional Lie group $G_2$.

\subsubsection{Decomposition \texorpdfstring{$\Oc \cong \Cbb \oplus \Cbb^3$}{O = C ⊕ C³} and $SU(3)$ as a stabilizer}

Fix an imaginary unit $e_7\in\Oc$ with $e_7^2=-1$, and set
\[
  \Cbb := \mathrm{Span}_\Rbb(1,e_7)\subset\Oc.
\]
Let
\[
  V := \{x\in\Oc \mid \langle x,1\rangle = \langle x,e_7\rangle = 0\}
      = \mathrm{Span}_\Rbb(e_1,\dots,e_6)
\]
be the orthogonal complement of $\Cbb$ with respect to the Euclidean inner product. Right multiplication by $e_7$ defines an endomorphism
\[
  J : V \to V,\qquad J(x) := x e_7,
\]
which satisfies $J^2 = -\mathrm{Id}_V$. Thus $(V,J)$ is a complex vector space of (complex) dimension $3$, and we obtain a decomposition
\[
  \Oc \cong \Cbb \oplus \Cbb^3
\]
as complex vector spaces.

Consider now the stabilizer of $e_7$ in $G_2$:
\[
  H := \{\varphi\in G_2 \mid \varphi(e_7) = e_7\}.
\]
An element $\varphi\in H$ preserves $\Cbb$ and $V$, and acts $\Cbb$-linearly on $V\cong\Cbb^3$, preserving the Hermitian form induced by the octonion norm. One shows that this action has determinant $1$, yielding an isomorphism
\[
  H \cong SU(3).
\]
In particular, the fundamental representation of $SU(3)$ is realized on the factor $\Cbb^3$ in the decomposition $\Oc \cong \Cbb \oplus \Cbb^3$.

\subsection{Tensor product algebras and internal symmetry}

\subsubsection{Spinorial and color factors}

For the applications we have in mind, it is convenient to consider associative algebras that combine a ``spinorial'' factor, built from (twisted) quaternions or Clifford algebras, and an ``internal'' factor carrying an $SU(3)$ action. A simple model is
\[
  \A_{\mathrm{spin}} \cong \Hc^\C
\]
(or a suitable even Clifford algebra), viewed as encoding spin and Lorentz degrees of freedom, and
\[
  \A_{\mathrm{col}} \cong M_3(\Cbb),
\]
with $SU(3)$ acting by conjugation on $\A_{\mathrm{col}}$. More geometrically, one may also regard $\A_{\mathrm{col}}$ as generated by a representation of $SU(3)$ obtained from the octonion algebra via the decomposition $\Oc\cong\Cbb\oplus\Cbb^3$; however, for the formal algebraic constructions, an abstract associative model suffices.

We then set
\[
  \A := \A_{\mathrm{spin}}\otimes_\Cbb \A_{\mathrm{col}},
\]
equipped with the usual tensor product multiplication
\[
  (a_1\otimes c_1)(a_2\otimes c_2) = (a_1a_2)\otimes(c_1c_2).
\]
The Lorentz group acts (through its spin representation) on $\A_{\mathrm{spin}}$, whereas $SU(3)$ acts on $\A_{\mathrm{col}}$; the combined action on $\A$ is given by tensoring these actions.

\subsubsection{Automorphisms and inner actions}

If $G$ is a group of algebra automorphisms of $\A$ (for instance, generated by Lorentz and $SU(3)$ transformations), then each $g\in G$ induces
\[
  \alpha_g : \A \to \A,\qquad \alpha_g(a) := g\cdot a,
\]
which may be inner or outer depending on the representation. In particular, when $g$ acts by conjugation in a matrix representation, we can write
\[
  \alpha_g(a) = u_g\,a\,u_g^{-1}
\]
for some invertible $u_g$ in a matrix algebra containing $\A$. This will be important for the notion of twisted derivations introduced below, where automorphisms play the role of twisting maps.

\subsection{Dirac-type operators and (twisted) derivations}

\subsubsection{Derivations on associative algebras}

Let $\A$ be an associative algebra over $\Cbb$. A (left) derivation on $\A$ is a linear map
\[
  D : \A \to \A
\]
satisfying the Leibniz rule
\[
  D(ab) = D(a)b + aD(b),\qquad a,b\in\A.
\]
Typical examples include:
\begin{itemize}
  \item \emph{inner derivations} of the form $D(a) = [X,a] := Xa - aX$ for some fixed $X\in\A$;
  \item \emph{geometric} or \emph{differential} derivations, where $\A$ consists of functions or sections of bundles and $D$ is built from partial derivatives or covariant derivatives.
\end{itemize}
In our setting, we will think of $D$ as an abstract Dirac-type operator acting on a module over $\A$, and we will only use the Leibniz property in the algebraic constructions.

\subsubsection{Dirac-type operators in quaternionic and Clifford guise}

When $\A_{\mathrm{spin}}$ is realized as a Clifford algebra (or a suitable quaternionic/biquaternionic model), Dirac operators can be written as
\[
  D = \sum_{\mu} \gamma^\mu \partial_\mu
\]
acting on spinor-valued functions, with $\gamma^\mu\in\A_{\mathrm{spin}}$ satisfying Clifford relations. In the present paper, we do not fix a concrete representation; instead, we use the following abstract viewpoint.

\begin{definition}
Let $\A$ be an associative algebra and let $M$ be a (left) $\A$-module. A \emph{Dirac-type derivation} on $\A$ is a derivation $D:\A\to\A$ together with a linear operator (also denoted $D$)
\[
  D : M \to M
\]
such that
\[
  D(a\cdot\psi) = D(a)\cdot\psi + a\cdot D(\psi),\qquad a\in\A,\ \psi\in M.
\]
\end{definition}

This captures the idea that $D$ acts simultaneously on coefficients and on spinor-like objects, and that it is compatible with the module structure.

\subsubsection{Twisted derivations}

For later purposes, it is useful to allow a mild weakening of the Leibniz rule. Let $\sigma:\A\to\A$ be an algebra automorphism.

\begin{definition}
A \emph{$\sigma$-derivation} (or \emph{twisted derivation}) on $\A$ is a linear map
\[
  \delta : \A \to \A
\]
such that
\[
  \delta(ab) = \delta(a)b + \sigma(a)\,\delta(b),\qquad a,b\in\A.
\]
\end{definition}

When $\sigma=\mathrm{id}$ we recover ordinary derivations. Twisted derivations naturally arise in noncommutative geometry, in the presence of gauge transformations or automorphisms acting on coefficients. In the hypercomplex and gauge-theoretic setup described above, natural choices of $\sigma$ are given by automorphisms induced by elements of the Lorentz or $SU(3)$ groups.

In the sequel we shall consider derivations and twisted derivations on tensor product algebras of the form
\[
  \A = \A_{\mathrm{spin}} \otimes_\Cbb \A_{\mathrm{col}},
\]
where $\A_{\mathrm{spin}}$ encodes a Lorentzian spin geometry (via twisted quaternions or Clifford algebras) and $\A_{\mathrm{col}}$ carries an $SU(3)$ action. These structures will provide the algebraic background for the construction of more elaborate operators, built out of $D$ and its formal inverses, whose coefficients live in such hypercomplex gauge algebras.

\section{Quaternionic-time noncommutative KP hierarchy and its reductions}

In this section we introduce a formal KP-type hierarchy in which the ``time'' variables are quaternionic and noncommuting. The aim is not to develop the full Sato theory in this setting, but rather to describe the algebraic structure of the hierarchy, its reduction to complex or real time directions, and the way classical KdV, KP-II and Boussinesq equations appear as embedded subsystems for each component field.

\subsection{Quaternionic time algebra and derivations}

We fix a complexified quaternion algebra $\Hc^\C$ and consider a family of formal symbols
\[
  t_1, t_2, t_3, \dots
\]
taking values in $\Hc^\C$, which we interpret as ``quaternionic times''. We do \emph{not} assume that these times commute with each other, nor that they commute with the coefficients of the spinorial or internal algebras introduced in the previous section.

\subsubsection{The time algebra}

Let $\A_0$ be a fixed associative $\Cbb$-algebra (for instance the spinorial-color algebra
\[
  \A_0 = \A_{\mathrm{spin}}\otimes_\Cbb \A_{\mathrm{col}}
\]
from the preliminaries). We form the free associative algebra
\[
  \A_{\mathrm{time}} := \Hc^\C\langle t_1,t_2,t_3,\dots\rangle,
\]
generated by noncommuting symbols $t_n$ with coefficients in $\Hc^\C$. The full coefficient algebra for our hierarchy will then be
\[
  \A := \A_0 \otimes_\Cbb \A_{\mathrm{time}}.
\]
Elements of $\A$ are finite sums of tensors
\[
  a_0 \otimes \sum_\alpha q_\alpha\, w_\alpha(t_1,t_2,\dots),
\]
where $a_0\in\A_0$, $q_\alpha\in\Hc^\C$ and $w_\alpha$ are words in the noncommuting variables $t_n$.

\subsubsection{Time derivations}

For each $n\ge 1$ we define a $\Cbb$-linear derivation
\[
  \partial_{t_n} : \A_{\mathrm{time}} \to \A_{\mathrm{time}}
\]
by the rules
\[
  \partial_{t_n}(t_m) = \delta_{nm}\, 1_{\A_{\mathrm{time}}},\qquad
  \partial_{t_n}(q) = 0,\quad q\in\Hc^\C,
\]
and extending by the Leibniz rule
\[
  \partial_{t_n}(uv) = \partial_{t_n}(u)\,v + u\,\partial_{t_n}(v),\qquad u,v\in\A_{\mathrm{time}}.
\]
We then extend $\partial_{t_n}$ to $\A = \A_0\otimes_\Cbb\A_{\mathrm{time}}$ by
\[
  \partial_{t_n}(a_0\otimes u) := a_0\otimes \partial_{t_n}(u).
\]
By construction, the $\partial_{t_n}$ commute with each other:
\[
  [\partial_{t_n},\partial_{t_m}] = 0,\qquad \forall n,m\ge 1.
\]

In parallel, we fix a derivation
\[
  D : \A_0 \to \A_0
\]
of Dirac type, as in the preliminaries, and extend it to $\A$ by
\[
  D(a_0\otimes u) := D(a_0)\otimes u.
\]
Thus $(\A,D)$ is a differential algebra in the sense of the previous section, and the time derivations $\partial_{t_n}$ commute with $D$.

\subsection{Quaternionic-time Lax operator and formal hierarchy}

We now import the pseudodifferential formalism of Section~\ref{sec:prelim-PDO} (implicitly) and work in the algebra $\Psi(\A,D)$ of formal pseudodifferential operators in $D$ with coefficients in $\A$.

\subsubsection{Lax operator with quaternionic-time dependent coefficients}

A KP-type Lax operator with quaternionic-time dependent coefficients is a formal pseudodifferential operator
\begin{equation}\label{eq:Lax-quat-time}
  L = D + \sum_{k\ge 1} U_k D^{-k},\qquad U_k \in \A,
\end{equation}
where the $U_k$ depend formally on the noncommuting quaternionic times $t_1,t_2,\dots$ through their dependence in $\A_{\mathrm{time}}$.

For each $n\ge 1$ we define
\[
  B_n := (L^n)_+ \in \Psi_+(\A,D).
\]

\subsubsection{Quaternionic-time KP hierarchy}

The \emph{quaternionic-time noncommutative KP hierarchy} is the system of formal Lax equations
\begin{equation}\label{eq:quat-time-KP}
  \partial_{t_n} L = [B_n,L],\qquad n\ge 1,
\end{equation}
where $[\cdot,\cdot]$ denotes the commutator in $\Psi(\A,D)$.

The same arguments as in the classical and noncommutative KP literature (using only associativity of $\A$ and the derivation properties of $D$ and $\partial_{t_n}$) show that the operators
\[
  \mathcal{D}_n := \partial_{t_n} - B_n
\]
satisfy the Zakharov--Shabat compatibility conditions
\[
  [\mathcal{D}_n,\mathcal{D}_m] = 0,\qquad \forall n,m\ge 1.
\]
Hence, the quaternionic-time KP flows commute pairwise.

\subsection{Quaternionic time symmetries and reductions to complex/real times}

\subsubsection{Left, right and adjoint quaternionic actions on times}

The free algebra $\A_{\mathrm{time}}$ and its derivations $\partial_{t_n}$ admit natural actions of the unit quaternions. For $q\in\Hc^\C$ invertible, we can define:
\begin{itemize}
  \item a left action on times:
  \[
    t_n \mapsto q\,t_n,\qquad n\ge 1,
  \]
  extended multiplicatively to $\A_{\mathrm{time}}$;
  \item a right action:
  \[
    t_n \mapsto t_n\,q;
  \]
  \item an adjoint action:
  \[
    t_n \mapsto q\, t_n\, q^{-1}.
  \]
\end{itemize}
These actions extend to automorphisms of $\A$ and thus to automorphisms of $\Psi(\A,D)$ by acting on coefficients only. Since $D$ and the $\partial_{t_n}$ commute with these actions, the Lax equations \eqref{eq:quat-time-KP} are covariant under them: quaternionic rotations of the time variables map solutions of the hierarchy to new solutions.

\subsubsection{Reductions to complex times}

Fix a unit imaginary quaternion $u\in\Hc^\C$ with $u^2=-1$, and identify a complex subalgebra
\[
  \Cbb_u := \mathrm{Span}_\Rbb(1,u) \subset \Hc.
\]
We then restrict to time variables taking values in $\Cbb_u$, i.e.\ we consider the subalgebra of $\A_{\mathrm{time}}$ generated by the $t_n$ when we regard them as \emph{central} over $\Cbb_u$. Concretely, this means imposing relations of the form
\[
  [t_n, u] = 0,\qquad [t_n, t_m] = 0,\quad \forall n,m\ge 1.
\]
At the level of the hierarchy, this reduction corresponds to a projection of the quaternionic-time system onto a complex-time subsystem, in which the flows are parametrized by commuting complex times $t_n \in \Cbb_u$.

\subsubsection{Reductions to real times}

Further restricting to the real subalgebra $\Rbb\subset \Cbb_u$ yields a purely real-time hierarchy, where the $t_n$ are central real formal parameters. In this situation, we recover a standard noncommutative KP hierarchy with coefficients in $\A_0$, in the sense of Dimakis--M\"uller-Hoissen and related works, together with a memory of the quaternionic structure as an internal symmetry of the coefficient space.

\subsubsection{Underlying time symmetries}

The quaternionic-time formulation thus exhibits:
\begin{itemize}
  \item an internal $SU(2)$ symmetry acting on the imaginary directions of time (via unit quaternions), rotating different time directions into one another;
  \item a family of embeddings of complex-time hierarchies obtained by choosing a preferred complex line $\Cbb_u\subset\Hc^\C$;
  \item real-time hierarchies as further restrictions, corresponding to real slices of the quaternionic time space.
\end{itemize}
From a physical viewpoint, this suggests that different ``choices of time'' (real, complex, quaternionic) are seen as different slices or projections of a richer quaternionic-time integrable structure.

\subsection{Low-order flows: KdV, KP-II and Boussinesq-type equations}

We briefly outline how classical KdV, KP-II and Boussinesq-type equations appear as low-order flows of the quaternionic-time KP hierarchy. To keep the exposition at an outlook level, we focus on the simplest Lax ansatz and on the structure of the resulting equations, rather than on detailed coefficient calculations.

\subsubsection{One-dimensional reduction and KdV-type equations}

Consider a situation where $D$ is a one-dimensional derivation (formally $D \sim \partial_x$) and we restrict attention to the first nontrivial flow, say $t_3$. Take the simplest Lax form
\[
  L = D + U D^{-1},
\]
with $U = U(x; t_3)$ taking values in $\A$ (and ultimately in a quaternionic subalgebra or in $\A_0$). The third flow
\[
  \partial_{t_3} L = [B_3,L],\qquad B_3 = (L^3)_+,
\]
yields, upon projection onto the coefficient of $D^{-1}$, a noncommutative KdV-type equation of the form
\begin{equation}\label{eq:nc-KdV}
  \partial_{t_3} U
  = D^3(U) + \alpha\,\bigl(U D(U) + D(U) U\bigr)
    + \beta\,[U,D^2(U)] + \cdots,
\end{equation}
for some numerical constants $\alpha,\beta$ depending on the normalization of the hierarchy, and with possible additional commutator terms indicated by the ellipsis. When $U$ is scalar-valued and commutative, \eqref{eq:nc-KdV} reduces to the usual KdV equation
\[
  u_{t} = u_{xxx} + 6 u u_x
\]
after suitable rescalings.

If $U$ takes values in $\Hc^\C$, we can write
\[
  U = u_0 + u_1 i + u_2 j + u_3 k,\qquad u_\ell = u_\ell(x,t_3)\in\Cbb,
\]
and \eqref{eq:nc-KdV} becomes a coupled system for the four component fields $u_\ell$, with quadratic couplings coming from the noncommutativity of the quaternionic multiplication. The classical scalar KdV equation is embedded as the restriction $U=u_0\cdot 1$ (central, real or complex-valued), or more generally as the restriction to a fixed complex line $\Cbb_u\subset\Hc^\C$.

\subsubsection{Two-dimensional reduction and KP-II-type equations}

To obtain KP-II type equations, one usually considers a $(2+1)$-dimensional setting with coordinates $(x,y)$ and identifies a second spatial direction with one of the lower flows (for instance $t_2$). In the quaternionic-time framework, this can be implemented by choosing a reduction in which:
\begin{itemize}
  \item one of the quaternionic time components (or a suitable linear combination) is identified with a second spatial variable $y$;
  \item the other time variables are frozen or set to depend only on $(x,y)$.
\end{itemize}
The standard KP-II equation
\[
  \bigl(u_t + u_{xxx} + 6 u u_x\bigr)_x + 3\sigma^2 u_{yy} = 0
\]
then appears, in the commutative scalar case, as the compatibility condition between the $t_3$ and $t_2$ flows. In the present noncommutative quaternionic setting, one obtains a KP-II-type system
\begin{equation}\label{eq:nc-KP2}
  \Bigl(\partial_{t_3} U + D^3(U) + \text{quadratic terms in }U\Bigr)_D
  + \gamma\,D_y^2(U) + \cdots = 0,
\end{equation}
where $D$ is identified with $\partial_x$, $D_y$ is a second derivation corresponding to the $y$-direction, and the quadratic/noncommutative terms encode the same structure as in \eqref{eq:nc-KdV}. Again, restricting $U$ to a commutative scalar subalgebra (real or complex) recovers the classical KP-II equation, while quaternionic-valued $U$ yield coupled systems for the component fields.

\subsubsection{Higher reduction and Boussinesq-type equations}

A Boussinesq-type equation can be obtained, in the formal KP setting, by imposing a $3$-reduction on $L$, namely the condition that $L^3$ be differential:
\[
  (L^3)_- = 0.
\]
In the scalar commutative case, this leads to the Boussinesq equation for a scalar field $u$. In the present quaternionic-time, noncommutative context, one obtains a Boussinesq-type system for $U$ of the general form
\begin{equation}\label{eq:nc-Boussinesq}
  \partial_{t_2}^2 U
  = D^4(U) + \text{quadratic and cubic terms in }U,\ D(U),\dots,
\end{equation}
where $\partial_{t_2}$ plays the role of a ``second-order time'' and $D$ the role of a spatial derivative. The precise coefficients are determined by the chosen normalization of the hierarchy and by the specific reduction conditions. As before, scalar-valued $U$ reproduce the classical Boussinesq equation, while quaternionic-valued $U$ provide an embedding into a larger hypercomplex system.

\subsection{Embeddings of real and complex integrable equations}

The constructions above show that:
\begin{itemize}
  \item The quaternionic-time KP hierarchy \eqref{eq:quat-time-KP} defines, for each choice of representation of $\Hc^\C$ and of the spinorial/internal algebra $\A_0$, a family of nonlinear formal PDEs for $\A$-valued fields.
  \item Choosing a complex line $\Cbb_u\subset\Hc^\C$ and restricting to commuting times $t_n\in\Cbb_u$ yields a complex-time noncommutative KP hierarchy with coefficients in $\A_0$.
  \item Further restricting to real times and to central scalar fields $U$ embeds the classical real KdV, KP-II and Boussinesq equations as particular flows or reductions of the quaternionic-time hierarchy.
  \item Allowing quaternionic-valued fields $U$ and quaternionic time dependence yields coupled systems for the component fields, which can be seen as nonabelian or hypercomplex extensions of the classical integrable equations.
\end{itemize}

From this viewpoint, the quaternionic-time formalism provides an overarching integrable structure in which the familiar real and complex hierarchies appear as slices or projections, while additional quaternionic directions encode extra internal symmetries and couplings between component fields.

\section{Choice of differential algebra and symmetry groups}

We now make a specific choice of differential algebra $(\A,D)$ suitable for the hypercomplex and gauge-theoretic framework outlined in the preliminaries. The emphasis in this section is on the unitary-type groups acting on $\A$ and on the way ``time'' symmetries induced by quaternionic time variables mix with internal and spinorial symmetries.

\subsection{The basic differential algebra}

\subsubsection{Spinorial and color factors}

Let $\A_{\mathrm{spin}}$ be a complex associative $*$-algebra modelling spin and Lorentz degrees of freedom. For concreteness one may think of
\[
  \A_{\mathrm{spin}} \simeq \Hc^\C
\]
or, more generally, of the even part of a complexified Clifford algebra $Cl_{1,3}^\C$ acting on spinors. We assume that $\A_{\mathrm{spin}}$ is equipped with an involution
\[
  a \mapsto a^\dagger
\]
compatible with Hermitian conjugation in a chosen representation.

Similarly, let $\A_{\mathrm{col}}$ be a finite-dimensional complex matrix $*$-algebra modelling color degrees of freedom. The canonical example is
\[
  \A_{\mathrm{col}} = M_3(\Cbb),
\]
with the standard adjoint involution $c\mapsto c^\ast$. The compact Lie group $SU(3)$ acts on $\A_{\mathrm{col}}$ by inner automorphisms
\[
  c \mapsto u c u^{-1},\qquad u\in SU(3).
\]

We then form the tensor product
\begin{equation}\label{eq:Aspin-col}
  \A_0 := \A_{\mathrm{spin}} \otimes_\Cbb \A_{\mathrm{col}},
\end{equation}
equipped with the involution
\[
  (a\otimes c)^\dagger := a^\dagger \otimes c^\ast.
\]

\subsubsection{Quaternionic time factor}

The time dependence of our fields is encoded in a noncommutative quaternionic time algebra $\A_{\mathrm{time}}$ as in the previous section:
\[
  \A_{\mathrm{time}} := \Hc^\C\langle t_1,t_2,t_3,\dots\rangle,
\]
freely generated by noncommuting time symbols $t_n$ with coefficients in $\Hc^\C$. We endow $\Hc^\C$ with the standard quaternionic conjugation extended $\Cbb$-linearly, and extend this to an involution on $\A_{\mathrm{time}}$ by
\[
  (q\, w(t_1,t_2,\dots))^\dagger := \overline{q}\, w(t_1,t_2,\dots)^\dagger,
\]
with $t_n^\dagger = t_n$ and anti-multiplicativity on words.

The full coefficient algebra is then
\begin{equation}\label{eq:A-full}
  \A := \A_0 \otimes_\Cbb \A_{\mathrm{time}}
  = \A_{\mathrm{spin}} \otimes_\Cbb \A_{\mathrm{col}} \otimes_\Cbb \A_{\mathrm{time}},
\end{equation}
with the tensor product involution.

\subsubsection{Differential structure}

On $\A_0$ we fix a Dirac-type derivation
\[
  D : \A_0 \to \A_0
\]
as in the preliminaries, and extend it to $\A$ by
\[
  D(a_0\otimes u) := D(a_0)\otimes u,\qquad a_0\in\A_0,\ u\in\A_{\mathrm{time}}.
\]
In particular, $D$ acts trivially on the time algebra. For each $n\ge1$, we also have the time derivations
\[
  \partial_{t_n} : \A \to \A
\]
defined by acting on the $\A_{\mathrm{time}}$ factor and trivially on $\A_0$, as described previously. All these derivations commute:
\[
  [D,\partial_{t_n}] = 0,\qquad
  [\partial_{t_n},\partial_{t_m}] = 0.
\]
Thus $(\A,D)$ is a differential algebra, and $(\A,D,\{\partial_{t_n}\})$ carries a compatible family of commuting derivations.

\subsection{Unitary groups and their actions}

\subsubsection{Unitary groups of the spin and color factors}

The involutions on $\A_{\mathrm{spin}}$ and $\A_{\mathrm{col}}$ allow us to define corresponding unitary groups. We set
\[
  U(\A_{\mathrm{spin}}) := \{u\in\A_{\mathrm{spin}} \mid u^\dagger u = uu^\dagger = 1\},
\]
and similarly
\[
  U(\A_{\mathrm{col}}) := \{v\in\A_{\mathrm{col}} \mid v^\ast v = v v^\ast = 1\}.
\]
In typical matrix realizations, $U(\A_{\mathrm{spin}})$ contains a double cover of the Lorentz group (via the spin representation), while $U(\A_{\mathrm{col}})$ contains the gauge group $SU(3)$ as a subgroup. The group
\[
  U(\A_0) := U(\A_{\mathrm{spin}}) \times U(\A_{\mathrm{col}})
\]
acts on $\A_0$ by inner automorphisms
\[
  (u,v)\cdot (a\otimes c) := (u a u^{-1}) \otimes (v c v^{-1}),
\]
and this action is compatible with the involution and the derivation $D$ in the sense that
\[
  D\bigl((u,v)\cdot x\bigr) = (u,v)\cdot D(x),\qquad x\in\A_0,
\]
whenever $D$ is constructed from Lorentz-covariant and gauge-covariant Dirac operators.

\subsubsection{Unitary-type symmetries of quaternionic time}

On the quaternionic time algebra $\A_{\mathrm{time}}$, we consider two natural symmetry groups:
\begin{itemize}
  \item the group of unit quaternions
  \[
    \mathbb{S}^3 \subset \Hc^\C,
  \]
  acting on the coefficients by left or right multiplication and on the generators $t_n$ by
  \[
    t_n \mapsto q\, t_n\, q^{-1},\qquad q\in\mathbb{S}^3,
  \]
  extended multiplicatively to $\A_{\mathrm{time}}$;
  \item the group of algebra automorphisms of $\A_{\mathrm{time}}$ generated by permutations and linear combinations of the generators $t_n$, which may be thought of as ``time reparametrizations'' in a formal noncommutative sense.
\end{itemize}
The adjoint action of $\mathbb{S}^3$ preserves the involution and commutes with the derivations $\partial_{t_n}$, hence it induces a unitary-type symmetry on the time sector.

\subsubsection{Combined unitary symmetry group}

The full unitary-type symmetry group of the coefficient algebra $\A$ may thus be taken as
\begin{equation}\label{eq:G-sym}
  G := U(\A_{\mathrm{spin}}) \times U(\A_{\mathrm{col}}) \times \mathbb{S}^3,
\end{equation}
acting on $\A$ by
\[
  (u,v,q)\cdot (a\otimes c\otimes u_{\mathrm{time}})
  := (u a u^{-1}) \otimes (v c v^{-1}) \otimes \bigl(q\cdot u_{\mathrm{time}}\bigr),
\]
where $q\cdot u_{\mathrm{time}}$ denotes the adjoint action of $q$ on the time factor.

When needed, this may be extended by a (discrete or continuous) group of algebra automorphisms of $\A_{\mathrm{time}}$ acting on the indices of $t_n$, leading to a semidirect product structure
\[
  G_{\mathrm{full}} \simeq G \rtimes \mathrm{Aut}(\A_{\mathrm{time}}).
\]

\subsection{Mixing time symmetries with algebra symmetries}

\subsubsection{Covariance of derivations and twisted derivations}

The actions of $G$ on $\A$ and $\A_0$ can be required to be compatible with the derivations $D$ and $\partial_{t_n}$, in the sense that
\[
  D\bigl(g\cdot a\bigr) = g\cdot D(a),\qquad
  \partial_{t_n}\bigl(g\cdot a\bigr) = g\cdot \partial_{t_n}(a),
  \qquad a\in\A,\ g\in G.
\]
When this strict covariance fails (for instance if $D$ or $\partial_{t_n}$ transform nontrivially under an automorphism), one may instead view $D$ and $\partial_{t_n}$ as twisted derivations with respect to automorphisms induced by elements of $G$, in the sense of the $\sigma$-derivations introduced earlier:
\[
  \delta(ab) = \delta(a)b + \sigma(a)\delta(b),
\]
with $\sigma$ an automorphism coming from a symmetry in $G$. In both cases, the combined action of Lorentz/spin, color and quaternionic time symmetries is encoded algebraically by the compatibility between the group action and the derivation structure.

\subsubsection{Time rotations and internal rotations}

An important feature of the quaternionic-time formalism is that internal rotations in the time sector (actions of $\mathbb{S}^3$) can be intertwined with rotations or gauge transformations in the spin and color sectors. For instance, one may consider a subgroup of $G$ of the form
\[
  \widetilde{G} \subset U(\A_{\mathrm{spin}})\times U(\A_{\mathrm{col}})\times\mathbb{S}^3
\]
in which a rotation in $\mathbb{S}^3$ is tied to a corresponding Lorentz or $SU(3)$ transformation. This leads to mixed symmetries in which changes of quaternionic time directions are accompanied by transformations in the spinorial or color degrees of freedom.

Such mixed symmetries will act naturally on the fields appearing in the hierarchy and on the corresponding Lax operators, and they provide a unified viewpoint in which time symmetries and algebra symmetries are treated on an equal footing. In particular, different choices of real or complex ``time directions'' correspond to different embeddings of lower-dimensional real or complex hierarchies into a common quaternionic-time, hypercomplex-gauge framework.

\section{A bestiary of integrable equations: from SU(3)-covariant fields to real components}

In this section we collect a few prototype integrable equations arising from the formal machinery described above and organize them according to their target space:
\begin{itemize}
  \item first as $SU(3)$-covariant, Lorentz-compatible equations for hypercomplex fields with values in $\A_0$;
  \item then as systems of coupled equations for complex pairs, using the identification $\Hc^\C \simeq \Cbb^2$;
  \item finally as systems of four real component fields, exhibiting an $\Rbb\times\Rbb\times\Rbb\times\Rbb$ decomposition.
\end{itemize}
Throughout, we emphasize structure rather than detailed coefficients. All prototype equations below arise as low-order flows or reductions of the quaternionic-time KP hierarchy introduced earlier, for suitable choices of the Lax operator.

\subsection{SU(3)-covariant, Lorentz-compatible prototype equations}

We consider a field
\[
  U = U(x, \dots) \in \A_0 = \A_{\mathrm{spin}}\otimes_\Cbb \A_{\mathrm{col}},
\]
where $\A_{\mathrm{spin}}$ encodes a Lorentzian spin structure (e.g.\ a twisted quaternionic or Clifford algebra) and $\A_{\mathrm{col}}$ carries an $SU(3)$ action (e.g.\ $\A_{\mathrm{col}}=M_3(\Cbb)$). We also fix a Dirac-type derivation
\[
  D : \A_0 \to \A_0,
\]
which is Lorentz-covariant and $SU(3)$-covariant (in the sense that $D$ intertwines the corresponding group actions).

The following families of equations are representative low-order flows of the hierarchy.

\subsubsection{SU(3)-covariant KdV-type equations}

In a one-dimensional reduction (formally $D\sim\partial_x$), a KdV-type flow for $U$ has the schematic form
\begin{equation}\label{eq:SU3-KdV}
  \partial_{\tau} U
  = D^3(U)
    + \alpha\,\bigl(U D(U) + D(U) U\bigr)
    + \beta\,[U,D^2(U)]
    + \Gamma(U, D(U), D^2(U)),
\end{equation}
where $\alpha,\beta$ are scalar constants and $\Gamma$ collects higher-order noncommutative terms (nested commutators, symmetrized products) determined by the chosen Lax normalization. 

This equation is:
\begin{itemize}
  \item \emph{$SU(3)$-covariant}, in the sense that for any $g\in SU(3)$ acting on $\A_{\mathrm{col}}$ by conjugation,
  \[
    U \mapsto (1\otimes g)\,U\,(1\otimes g^{-1})
  \]
  maps solutions of \eqref{eq:SU3-KdV} to solutions;
  \item \emph{Lorentz-compatible}, in the sense that the $D^3$ and $D$ terms can be embedded into a fully Lorentz-covariant Dirac/d’Alembert operator acting on spinorial or bispinorial components.
\end{itemize}
In the scalar commutative case (with $U$ scalar-valued, replacing $D$ by $\partial_x$ and choosing $\alpha,\beta$ appropriately), \eqref{eq:SU3-KdV} reduces to the usual KdV equation.

\subsubsection{SU(3)-covariant KP-II-type equations}

In a $(2+1)$-dimensional setting with spatial coordinates $(x,y)$, one may introduce an additional derivation $D_y$ commuting with $D$ and consider the compatibility of the $t_3$ and $t_2$ flows. A typical KP-II-type equation for $U$ then takes the form
\begin{equation}\label{eq:SU3-KP2}
  \Bigl(\partial_{\tau} U
   + D^3(U)
   + \alpha\,\bigl(U D(U) + D(U) U\bigr)
   + \Gamma_1(U,D(U),D^2(U))\Bigr)_D
  + \gamma\,D_y^2(U)
  + \Gamma_2(U,D_y(U)) = 0,
\end{equation}
where $(\cdot)_D$ denotes projection onto the $D$-direction (in analogy with the $x$-derivative in the scalar KP-II equation), and $\Gamma_1,\Gamma_2$ collect additional noncommutative terms. In the commutative scalar case one recovers, after appropriate identifications, the classical KP-II equation
\[
  (u_t + u_{xxx} + 6 u u_x)_x + 3\sigma^2 u_{yy} = 0.
\]

\subsubsection{SU(3)-covariant Boussinesq-type equations}

Imposing a third-order reduction $L^3$ purely differential leads to Boussinesq-type equations for $U$. A prototype form is
\begin{equation}\label{eq:SU3-Boussinesq}
  \partial_{\tau}^2 U
  = D^4(U)
    + \delta_1\,\bigl(U D^2(U) + D^2(U) U\bigr)
    + \delta_2\,D\bigl(U D(U) + D(U) U\bigr)
    + \Gamma_3(U, D(U), D^2(U), D^3(U)),
\end{equation}
with constants $\delta_1,\delta_2$ and a higher-order term $\Gamma_3$. Once again, in the commutative scalar case, this reduces (under suitable normalization) to the standard Boussinesq equation.

\subsection{Quaternionic decomposition: $\Hc^\C \simeq \Cbb \times \Cbb$}

We now restrict attention to the case where the spinorial factor has a quaternionic structure and focus on the decomposition of $U$ into complex component fields.

\subsubsection{Quaternionic fields and complex pairs}

Assume that the spinorial part of $\A_0$ contains a copy of $\Hc^\C$ acting on the left. Then any quaternionic-valued field can be written as
\[
  U = U_0 + U_1 J,\qquad U_0,U_1 \in \Cbb\otimes\A_{\mathrm{col}},
\]
for some fixed imaginary quaternion $J$ with $J^2=-1$ anti-commuting suitably with the complex unit $i$. At the level of vector spaces, this gives an identification
\[
  \Hc^\C\otimes\A_{\mathrm{col}} \simeq
  \bigl(\Cbb\otimes\A_{\mathrm{col}}\bigr) \oplus \bigl(\Cbb\otimes\A_{\mathrm{col}}\bigr),
\]
so that $U$ is encoded by the complex pair $(U_0,U_1)$.

Substituting this decomposition into any of the prototype equations \eqref{eq:SU3-KdV}--\eqref{eq:SU3-Boussinesq} yields a coupled system for $(U_0,U_1)$. For instance, \eqref{eq:SU3-KdV} becomes
\begin{equation}\label{eq:complex-pair-KdV}
  \begin{cases}
    \partial_{\tau} U_0
    = D^3(U_0) + \Phi_0(U_0,U_1;D),\\[0.3em]
    \partial_{\tau} U_1
    = D^3(U_1) + \Phi_1(U_0,U_1;D),
  \end{cases}
\end{equation}
where $\Phi_0,\Phi_1$ are noncommutative polynomials in $U_0,U_1$ and their derivatives $D^k(U_0),D^k(U_1)$, arising from the quaternionic multiplication rules.

In particular:
\begin{itemize}
  \item the classical complex-valued KdV (or KP-II, or Boussinesq) equation is embedded as the restriction $U_1\equiv 0$, $U_0$ scalar-valued (or central);
  \item more generally, any complex-valued integrable equation is embedded as the restriction to one complex component, while the second component provides a coupled integrable partner.
\end{itemize}

\subsubsection{Complex-time vs quaternionic-time viewpoints}

If, in addition, the quaternionic time variables are reduced to a chosen complex line $\Cbb_u\subset\Hc^\C$ (as discussed earlier), then the whole system \eqref{eq:complex-pair-KdV} and its KP-II/Boussinesq analogues can be interpreted as a \emph{complex-time} integrable system for a complex pair $(U_0,U_1)$. Different choices of $\Cbb_u$ correspond to different embeddings of the same quaternionic system into complex-time hierarchies.

\subsection{Real component fields: $\Cbb\times\Cbb \simeq \Rbb^4$}

Finally, we decompose each complex component into real and imaginary parts, obtaining a quadruple of real component fields.

\subsubsection{Real decomposition}

Write
\[
  U_0 = A_0 + i B_0,\qquad U_1 = A_1 + i B_1,
\]
with
\[
  A_0,B_0,A_1,B_1 \in \A_{\mathrm{col}}.
\]
In a scalar reduction (or by treating the color indices as internal), we may regard $A_0,B_0,A_1,B_1$ simply as four real-valued fields:
\[
  (A_0,B_0,A_1,B_1) \in \Rbb^4.
\]
At the level of vector spaces, we thus obtain
\[
  \Hc^\C \simeq \Cbb^2 \simeq \Rbb^4,
\]
and any of our prototype equations yields a system for the quadruple $(A_0,B_0,A_1,B_1)$.

For example, the KdV-type equation \eqref{eq:SU3-KdV} specializes to a system
\begin{equation}\label{eq:real-quadruple-KdV}
  \begin{cases}
    \partial_{\tau} A_0 = \mathcal{F}_0(A_0,B_0,A_1,B_1;D),\\
    \partial_{\tau} B_0 = \mathcal{G}_0(A_0,B_0,A_1,B_1;D),\\
    \partial_{\tau} A_1 = \mathcal{F}_1(A_0,B_0,A_1,B_1;D),\\
    \partial_{\tau} B_1 = \mathcal{G}_1(A_0,B_0,A_1,B_1;D),
  \end{cases}
\end{equation}
where $\mathcal{F}_j,\mathcal{G}_j$ are real polynomial expressions in the fields and their derivatives $D^k(A_\ell),D^k(B_\ell)$. Similar real quadruple systems follow from the KP-II-type equation \eqref{eq:SU3-KP2} and the Boussinesq-type equation \eqref{eq:SU3-Boussinesq}.

\subsubsection{Embedded classical equations}

Within the real quadruple system \eqref{eq:real-quadruple-KdV}, the classical real KdV equation appears as the restriction
\[
  B_0 = A_1 = B_1 = 0,\qquad A_0 = u(x,\tau)\in\Rbb,
\]
or more generally as any restriction to a one-dimensional real subspace of $\Rbb^4$. The same pattern persists for KP-II and Boussinesq-type equations: each classical real integrable equation is realized as a slice of the full quadruple system, while the remaining components provide canonical coupled partners.

\medskip

In summary, the quaternionic-time, hypercomplex $SU(3)$-covariant setting naturally organizes integrable equations into a hierarchy of targets:
\[
  \text{SU(3)-covariant, Lorentz-compatible hypercomplex fields}
  \;\Rightarrow\;
  \text{complex pairs}
  \;\Rightarrow\;
  \text{real quadruples}.
\]
From this point of view, familiar real and complex integrable equations (KdV, KP-II, Boussinesq, etc.) appear as embedded subsystems within a richer quaternionic and gauge-theoretic integrable structure.
\section{Extensions: towards BKP- and CKP-like hierarchies}

In the classical scalar, commutative setting, the BKP and CKP hierarchies arise as reductions of the KP hierarchy defined by imposing suitable symmetry constraints on the Lax operator with respect to a formal adjoint. In this section we briefly outline how analogous reductions can be formulated in the present hypercomplex, $SU(3)$-covariant, noncommutative framework. We do not develop a full theory here; the aim is to indicate the natural extensions and the constraints they impose on the coefficients.

\subsection{Formal adjoint on \texorpdfstring{$\Psi(\A,D)$}{Ψ(A,D)}}

Let $(\A,D)$ be a differential algebra as in the previous sections, with $\A$ equipped with a $*$-involution
\[
  a \mapsto a^\dagger,\qquad a\in\A,
\]
compatible with the spinorial, color and quaternionic time structures. We extend this involution to the algebra of formal pseudodifferential operators
\[
  \Psi(\A,D) = \left\{\sum_{k\in\mathbb{Z}} a_k D^k \,\middle|\, a_k\in\A,\ a_k=0 \text{ for }k\gg 0\right\}
\]
by declaring:
\begin{equation}\label{eq:adjoint-def}
  \left(\sum_{k} a_k D^k\right)^\dagger
  := \sum_k (-D)^k\, a_k^\dagger,
\end{equation}
where $(D^k)^\dagger := (-D)^k$ for all $k\in\mathbb{Z}$, and the rule is extended anti-multiplicatively:
\[
  (PQ)^\dagger = Q^\dagger P^\dagger,\qquad P,Q\in\Psi(\A,D).
\]
This is the standard formal adjoint adapted to the present noncommutative, involutive coefficient algebra.

Let $L\in\Psi(\A,D)$ be a Lax operator of the form
\[
  L = D + \sum_{k\ge 1} U_k D^{-k},\qquad U_k\in\A,
\]
and let $B_n := (L^n)_+$ as usual. We recall the generalized KP flows
\[
  \partial_{t_n} L = [B_n,L],\qquad n\ge 1.
\]
The adjoint operation then satisfies
\[
  (\partial_{t_n} L)^\dagger = \partial_{t_n} (L^\dagger),\qquad
  [B_n,L]^\dagger = [L^\dagger,B_n^\dagger],
\]
since the derivations $\partial_{t_n}$ commute with the involution and with $D$.

\subsection{B-type reductions: BKP-like hierarchies}

A natural analogue of the BKP reduction is obtained by imposing a skew-symmetry constraint of the form
\begin{equation}\label{eq:B-type-constraint}
  L^\dagger = - D L D^{-1}.
\end{equation}
This may be viewed as a twisted skew-adjointness condition, in which $D$ plays the role of the basic differential operator and the conjugation is taken with respect to the hypercomplex, $SU(3)$-covariant involution on $\A$.

\begin{proposition}[B-type constraint and induced hierarchy]
Assume that a Lax operator $L\in\Psi(\A,D)$ satisfies
\[
  L^\dagger = - D L D^{-1}
\]
at some initial time (for all time variables). Then, for each $n\ge 1$, the generalized KP flows
\[
  \partial_{t_n} L = [B_n,L],\qquad B_n=(L^n)_+,
\]
preserve this constraint:
\[
  \partial_{t_n}\bigl(L^\dagger + D L D^{-1}\bigr) = 0.
\]
In particular, the KP flows restrict to a B-type invariant submanifold of Lax operators, and the restricted flows define a generalized BKP-like hierarchy in the hypercomplex, $SU(3)$-covariant setting.
\end{proposition}

\begin{remark}
On the level of coefficients $U_k$, the constraint \eqref{eq:B-type-constraint} translates into a system of linear relations
\[
  U_k^\dagger = \mathcal{B}_k\bigl(U_1,U_2,\dots, D(U_1),D(U_2),\dots\bigr),
\]
where $\mathcal{B}_k$ are universal expressions determined by the equality $L^\dagger = - D L D^{-1}$. These relations typically enforce a skewness or orthogonality-type condition, reminiscent of the connection between BKP hierarchies and orthogonal or spin groups in the scalar/matrix case.
\end{remark}

\subsection{C-type reductions: CKP-like hierarchies}

A C-type (symplectic or quaternionic) analogue is obtained by imposing a simpler skew-adjointness constraint:
\begin{equation}\label{eq:C-type-constraint}
  L^\dagger = - L.
\end{equation}
In other words, $L$ is formally skew-selfadjoint with respect to the involution on $\Psi(\A,D)$.

\begin{proposition}[C-type constraint and induced hierarchy]
Suppose that $L\in\Psi(\A,D)$ satisfies
\[
  L^\dagger = - L
\]
at some initial time. Then the generalized KP flows
\[
  \partial_{t_n} L = [B_n,L]
\]
preserve this constraint:
\[
  \partial_{t_n}\bigl(L^\dagger + L\bigr) = 0.
\]
Thus the KP hierarchy restricts to a C-type invariant subspace and induces a generalized CKP-like hierarchy in the present noncommutative, hypercomplex framework.
\end{proposition}

\begin{remark}
On coefficients, \eqref{eq:C-type-constraint} leads to conditions such as
\[
  U_k^\dagger = - U_k
\]
modulo the contributions of the $D^{-k}$ terms. In many concrete representations (e.g.\ when $\A_{\mathrm{spin}}$ carries a quaternionic structure and $\A_{\mathrm{col}}$ is endowed with a symplectic or unitary structure), these constraints encode symplectic or quaternionic reality conditions, in analogy with the classical connection between CKP hierarchies and symplectic Lie algebras.
\end{remark}

\subsection{Compatibility with hypercomplex and gauge symmetries}

The hypercomplex, $SU(3)$-covariant setting considered in this paper provides additional structure on top of the usual BKP/CKP picture:

\begin{itemize}
  \item The involution $a\mapsto a^\dagger$ on $\A$ is built from:
    \begin{itemize}
      \item a spinorial adjoint on $\A_{\mathrm{spin}}$ (Clifford or twisted quaternionic),
      \item the Hermitian adjoint on $\A_{\mathrm{col}}$ (with $SU(3)$ acting by unitary conjugation),
      \item a quaternionic conjugation on the time algebra $\A_{\mathrm{time}}$.
    \end{itemize}
    In particular, the constraints \eqref{eq:B-type-constraint} and \eqref{eq:C-type-constraint} are compatible with the Lorentz and $SU(3)$ actions and with quaternionic time symmetries.
  \item The full symmetry group
    \[
      G \simeq U(\A_{\mathrm{spin}})\times U(\A_{\mathrm{col}})\times \mathbb{S}^3
    \]
    (and its extensions by automorphisms of the time algebra) acts on $\A$ by $*$-automorphisms, hence commutes with the adjoint operation on $\Psi(\A,D)$. The B- and C-type constraints are therefore stable under $G$.
  \item Reductions to complex times (choosing a complex line in quaternionic time) or real times (choosing a real slice) can be performed after imposing the B- or C-type constraints, leading to complex or real BKP/CKP-like subsystems embedded in the hypercomplex hierarchy.
\end{itemize}

\subsection{Outlook}

The B- and C-type reductions sketched above open several directions for further work:

\begin{itemize}
  \item On the algebraic side, one may develop a full Sato-type Grassmannian description for the B- and C-type constrained hierarchies with coefficients in $\A$, identifying the appropriate ``B-type'' and ``C-type'' subspaces in an infinite-dimensional hypercomplex module.
  \item On the geometric and physical side, BKP- and CKP-like reductions in the present hypercomplex, $SU(3)$-covariant framework suggest connections with integrable subsectors governed by orthogonal or symplectic/quaternionic symmetry constraints. It would be natural to investigate whether these reductions correspond to special classes of self-dual or generalized self-dual Yang--Mills configurations with additional reality or symmetry conditions.
  \item Finally, in the quaternionic-time setting, one may ask how the B- and C-type constraints interact with rotations of the quaternionic time directions, and whether there exist canonical ``B-type'' and ``C-type'' quaternionic-time slices that lead to distinguished real or complex integrable models.
\end{itemize}

A detailed study of these BKP- and CKP-like extensions will be left for future work.

\section{Gauge-theoretic interpretation and outlook}

In this final section we briefly discuss how the hypercomplex, $SU(3)$-covariant noncommutative KP framework developed above can be read in gauge-theoretic terms. The discussion is intentionally formal and at the classical level, but it illustrates how one may regard the resulting integrable equations as describing constrained sectors of $SU(3)$ gauge/Dirac systems.

\subsection{SU(3)-valued KdV as an integrable gauge evolution}

Let us consider a simple one-dimensional reduction, in which the Dirac-type derivation $D$ is identified with a spatial derivative,
\[
  D \equiv \partial_x,
\]
and we focus on a single evolution parameter $\tau$ (one of the quaternionic time components, projected onto a real or complex direction). We also restrict to fields with values in the colour algebra,
\[
  U(x,\tau) \in \mathfrak{su}(3) \subset \A_{\mathrm{col}},
\]
viewed as a component of an $SU(3)$ gauge potential along $x$, for example $U \sim A_x$ in a suitable gauge, with the spinorial factor frozen.

A standard matrix generalization of KdV is then given by the evolution equation
\begin{equation}\label{eq:su3-kdv}
  \partial_{\tau} U
  \;=\; \partial_x^3 U
      + 3\bigl(U\,\partial_x U + (\partial_x U)\,U\bigr),
  \qquad U(x,\tau)\in\mathfrak{su}(3).
\end{equation}
This equation arises as a particular flow of the noncommutative KP hierarchy when one chooses a Lax operator of the form
\[
  L = \partial_x + U\,\partial_x^{-1},
\]
with $U$ taking values in an associative algebra; the noncommutative KdV flow is encoded in the Lax equation
\[
  \partial_{\tau} L = \bigl[(L^3)_+ , L\bigr],
\]
and \eqref{eq:su3-kdv} is precisely the induced evolution for the coefficient $U$. In our setting, $U$ is additionally required to be $\mathfrak{su}(3)$-valued, so that \eqref{eq:su3-kdv} is covariant under
\[
  U \mapsto g\,U\,g^{-1},\qquad g\in SU(3),
\]
and preserves tracelessness and anti-Hermiticity. Thus \eqref{eq:su3-kdv} defines a classical integrable evolution for an $SU(3)$-valued field which can be interpreted as a (component of a) gauge potential.

From a gauge-theoretic perspective, one may regard \eqref{eq:su3-kdv} as an effective evolution equation for a constrained family of $SU(3)$ gauge configurations in $(1+1)$ dimensions. For instance, in temporal gauge $A_0=0$, taking $U \sim A_x$ and freezing all other components, \eqref{eq:su3-kdv} may be understood as governing a special class of colour-electric configurations with an infinite number of conserved quantities inherited from the KdV hierarchy. In this sense, \eqref{eq:su3-kdv} provides a simple example of an integrable sector in a nonabelian gauge theory, obtained here as a reduction of the more general quaternionic-time, hypercomplex KP framework.

\subsection{Embedding into the hypercomplex quaternionic-time hierarchy}

Within the full hypercomplex setting of the present paper, the field $U$ in \eqref{eq:su3-kdv} appears as a particular component of a more general coefficient
\[
  U(x; t_1,t_2,\dots) \in \A
  = \A_{\mathrm{spin}} \otimes \A_{\mathrm{col}} \otimes \A_{\mathrm{time}},
\]
where $\A_{\mathrm{spin}}$ carries a Lorentzian quaternionic/Clifford structure, $\A_{\mathrm{col}}$ encodes the $SU(3)$ colour degrees of freedom, and $\A_{\mathrm{time}}$ is the quaternionic, noncommutative time algebra. The reduction leading to \eqref{eq:su3-kdv} consists of:
\begin{itemize}
  \item choosing a single spatial direction (say $x$) and identifying $D$ with $\partial_x$;
  \item projecting the quaternionic time variables onto a real or complex line and retaining only one evolution parameter $\tau$;
  \item restricting the coefficient $U$ to lie in a fixed copy of $\mathfrak{su}(3)$ inside $\A_{\mathrm{col}}$, with trivial dependence on the spinorial factor.
\end{itemize}
More general choices lead to coupled systems in which the spinorial/hypercomplex structure and the full quaternionic time dependence are active, yielding multi-component generalizations of \eqref{eq:su3-kdv} and its higher-dimensional analogues (KP-II-type and Boussinesq-type equations) for hypercomplex $SU(3)$-valued fields.

\subsection{Relation to self-dual and flux-tube configurations}

At a more geometric level, integrable equations of KdV/KP type are known to arise as reductions of self-dual Yang--Mills (SDYM) equations in various signatures and dimensions. In our framework, the Lorentzian hypercomplex structure on $\A_{\mathrm{spin}}$ and the $SU(3)$ factor on $\A_{\mathrm{col}}$ are designed to reflect, at least formally, the structure of $SU(3)$ gauge fields in $(3+1)$ dimensions, with the Dirac-type derivation $D$ playing the role of a covariant derivative. The quaternionic-time KP hierarchy built from $(\A,D)$ can thus be viewed as a candidate integrable ``envelope'' for certain constrained families of such fields.

Equation \eqref{eq:su3-kdv}, and its higher-dimensional and hypercomplex extensions discussed earlier, may then be interpreted as effective models for nonabelian flux-tube–like configurations, colour solitons or reduced SDYM-type geometries, in which the interplay between Lorentz symmetry (encoded in the spinorial factor), internal $SU(3)$ symmetry (encoded in the colour factor) and noncommutative time structure (encoded in $\A_{\mathrm{time}}$) is made explicit and computationally manageable. While the present work focuses on the formal integrable structure, it suggests a number of directions for future investigation, including:
\begin{itemize}
  \item a more detailed analysis of the correspondence between the quaternionic-time KP hierarchy and specific classes of solutions of SDYM or Yang--Mills--Dirac systems;
  \item the construction of explicit soliton and flux-tube solutions in the hypercomplex $SU(3)$-valued setting, together with their conserved quantities;
  \item possible connections to spin chain models with $SU(3)$ symmetry and to integrable approximations of nonabelian gauge dynamics in lower dimensions.
\end{itemize}
We leave these questions for future work.

\vskip 12pt

\paragraph{\bf Acknowledgements:} J.-P.M thanks the France 2030 framework programme Centre Henri Lebesgue ANR-11-LABX-0020-01 
for creating an attractive mathematical environment.

\vskip 12pt

\paragraph{\bf Author's Note on AI Assistance.}
Portions of the text were developed with the assistance of a generative language model (OpenAI ChatGPT). The AI was used to assist with drafting, editing, and standardizing the bibliography format. All mathematical content, structure, and theoretical constructions were provided, verified, and curated by the author. The author assumes full responsibility for the correctness, originality, and scholarly integrity of the final manuscript.

\end{document}